\documentstyle[prl,aps,epsfig,twocolumn]{revtex}

\begin{document}
\draft

\twocolumn[\hsize\textwidth\columnwidth\hsize\csname@twocolumnfalse\endcsname
\title{Two photon interference and optical free induction decay}
\author{P.Zhang and C. P. Sun $^{a,b}$}
\address{Institute of Theoretical Physics, the Chinese Academy of Science,
 Beijing, 100080, China}
\maketitle
\begin{abstract}
The two photon interference phenomenon is theoretically
investigated for the general situations with an arbitrary input
two photon state with and without photon polarization. For the
case without polarization, the necessary-sufficient condition for
the destructive interference of coincidence counting is given as
the symmetric pairing of photons in the light pulses. For both
case it is shown that the "dip" in coincidence curve can be
understood in terms of the free induction decay mechanism. This
observation predicts the destructive interference phenomenon to
occur even for certain cases with separable input two photon
state, but it can only be explained in terms of "the two photon
(not two photons )interference ".
\end{abstract}
 ]
In last two decades one of the most important progresses in quantum optics
is the experimental demonstration of the Einstern-Podolsky-Rosen (EPR) [1]
effect with the entangled photons. The creation of EPR entangled photons in
a spontaneous parametric down conversion (SPDC [2]) not only implements the
best test of Bell's inequality [3] with fewer loopholes [4-6], but also lays
foundations for some quantum information techniques [7]. This important
progress has inspired people to reconsider the profound observation that
\textquotedblleft ...photon... only interferes with itself.
\textquotedblright stated by Dirac in his famous book \textquotedblleft
\textit{The Principle of Quantum Mechanics \textquotedblright }[8].

Some interesting experiments of typical two photon interferometer seem to
illustrate the existence of interference between two different photons since
a curve of ``dip''was observed in the rate of two photon coincidence[9-11].
However, a series of refined experimental setups [12-14] concluded that
Dirac is correct. They argue that ``a two photon (not two photons) can also
only interfere with itself''\ [14] to account for the exotic interference
phenomenon. In this argument the crucial conception is the two photon or
bi-photon, the inseparable photon pair depicted by a EPR state. In fact, the
naive idea of ``destructive interference between the idler and signal
photons''\ can not give a correct prediction for a two photon interference
experiment ``with three arms''.

If one believe that the idea of bi-photon is indeed necessary for all the
two photon interference experiments, then two natural questions follow
immediately : 1. Does there exist the separable pair of photons to give the
destructive interference in the two photon coincidence counting rate? In
this case the separable pair of photons can be implemented experimentally in
the two independent pulses of photons. 2.If there is indeed a destructive
interference for two independent pulses of photons, does it imply
\textquotedblleft interference between the idler and signal
photons\textquotedblright ?

\bigskip

\section{Two photon interference of Type I}

In this paper the two questions will be answered in an universal framework
by considering a general input two photon state
\begin{equation}
|\Psi \rangle =\int_{0}^{\infty }d\omega \int_{0}^{\infty }d\tilde{\omega}%
f(\omega ,\tilde{\omega})a_{A}^{+}(\omega )a_{B}^{+}(\tilde{\omega}%
)|0\rangle .
\end{equation}%
The schematic setup for this consideration is illustrated in figure 1:

%
\begin{figure}[h]
\begin{center}
\includegraphics[width=7cm,height=4cm]{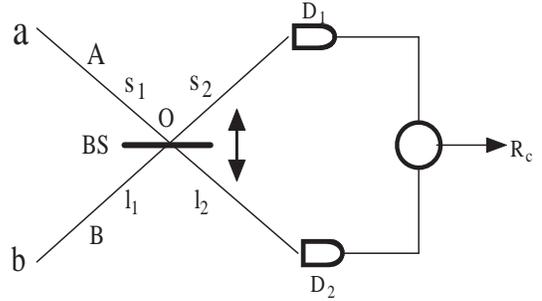}
\end{center}
\caption{Schematic of the two-photon interferometer}
\end{figure}
the two lights coming from points $a$ and $b$ are mixed at point $O$ on the
50-50 beam splitter $BS$ and detected by the two photon counters at $D_{1}$
and $D_{2}$. Correspondingly, $a_{A}^{+}(\omega )$ ( $a_{B}^{+}(\tilde{\omega%
})$ ) is the creation operator for the photon of frequency $\omega $ ( $%
\tilde{\omega}$ ) in the optical path mode $A$ ($B)$ from the point $a$ $(b)$
to $O$; $f(\omega ,\tilde{\omega})$ is the distribution function. The
optical path modes can be regarded as the modes of idler and signal lights
in the usual SPDC experiment with a special distribution function $f(\omega ,%
\tilde{\omega})\propto $ $\delta (\omega +\tilde{\omega}-\omega _{0})$
for a finite real number $\omega _{0}.$ The separable pair of photons
corresponds to the case with the factorized distribution function $f(\omega ,%
\tilde{\omega})=f_{1}(\omega )f_{2}(\tilde{\omega}).$

\begin{figure}[h]
\begin{center}
\includegraphics[width=5cm,height=5cm]{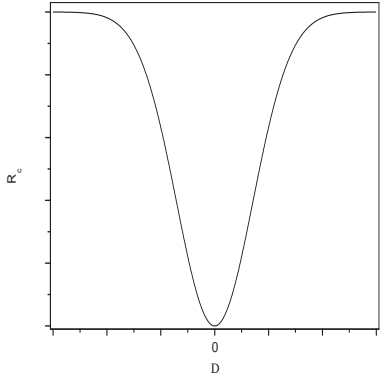}
\end{center}
\caption{The destructive interference}
\end{figure}

With this configuration of \textit{gedankenexperiment}, the above two
questions can be asked in an unique way : what kind of two photon state (or
what kind of distribution function $f(\omega ,\tilde{\omega})$) can result
in the curve of ``dip''\ of the rate of two photon coincidence, which is
illustrated in figures 2 as the typical destructive interference phenomenon.
Here, $D=l_{1}-s_{1}$ is the optical path difference between $aO$ and $bO;$ $%
aO=$ $s_{1},bO=$ $l_{1},$ $D_{1}O=s_{2},$ $D_{2}O=l_{2}$ are the lengths of
optical paths. The rate $R_{c}$ of coincident detection is measured as a
function of the position of $BS$ and there is a ''null''\ in coincidence
i.e., $R_{c}=0$ at $D=0$. This indicates a destructive interference. To
explain the anti-exponential decaying behavior beside the ''dip''\ (or
anti-peak) point, we will resort to the free induction decay mechanism [15].

In the following discussion we do not make assumptions about the light
source, which may be the entangled photon pair created by the BBO nonlinear
crystal or any two independent photon input pulse. We denote the positive
frequency parts of the electric field at detectors $D_{1}$ and $D_{2}$ by $%
E_{1}^{+}$ and $E_{2}^{+}$. We will choose a proper position as the
coordinate origin so that we could compute the coincidence rate
conveniently. According to Glauber's coherence theory [16,17], the
probability per unit (time)$^{2}$ that one photon is recorded at $D_{1}$ at
time $t_{1}$ and another at $D_{2}$ at time $t_{2}$ is just the second order
coherence function.%
\begin{equation}
G^{[2]}(t_{1},t_{2})=\langle \Psi
|E_{2}^{-}(t_{2})E_{1}^{-}(t_{1})E_{1}^{+}(t_{1})E_{2}^{+}(t_{2})|\Psi
\rangle
\end{equation}%
We recall that, when one consider the first order coherence the
superposition of two obvious ''paths'' is available. To describe the two
particle interference phenomenon due to the second order coherence in a
similar way, the generalized ``path''\ is introduced by considering $%
G^{[2]}(t_{1},t_{2})=|\Psi (t_{1},t_{2})|^{2}$ where the bi-photon wave
packet $\Psi (t_{1},t_{2})=\langle 0|E_{1}^{+}(t_{1})E_{2}^{+}(t_{2})|\Psi
\rangle $ is invoked as a two photon effective wave-function [14]. Most
recently this result was generalized for higher order coherence in
time-domain [18]. Then, the two photon coincidence rate can be given by the
two time integral $R_{c}=\int_{-\infty }^{\infty }dt_{1}dt_{2}|\Psi
(t_{1},t_{2})|^{2}$.

In present discussion, in terms of the positive frequency part of the $A$ ($%
B $) mode electric field at point $a$ ($b$) with the spectral density $%
g(\omega )$
\begin{equation}
E_{a(b)}^{+}(\tau )=\int_{0}^{\infty }d\omega g(\omega )a_{A(B)}(\omega
)e^{-i\omega \tau }
\end{equation}%
the local field operators
\begin{eqnarray}
E_{1}^{+}(t_{1}) &=&\frac{1}{\sqrt{2}}[iE_{a}^{+}(\tau _{1})+E_{b}^{+}(\tau
_{1}^{\prime })] \\
E_{2}^{+}(t_{2}) &=&\frac{1}{\sqrt{2}}[E_{a}^{+}(\tau _{2}^{\prime
})+iE_{b}^{+}(\tau _{2})]  \nonumber
\end{eqnarray}%
are implemented by the $50-50$ beam splitter with
\begin{eqnarray*}
\tau _{1} &=&t_{1}-\frac{s_{1}+s_{2}}{c},\tau _{2}=t_{2}-\frac{l_{1}+l_{2}}{c%
} \\
\tau _{1}^{\prime } &=&t_{1}-\frac{l_{1}+s_{2}}{c},\tau _{2}^{\prime }=t_{2}-%
\frac{s_{1}+l_{2}}{c}
\end{eqnarray*}%
where $c$ is the velocity of light. Then the two-photon wave function can be
written explicitly as%
\begin{equation}
\Psi =\int_{0}^{\infty }F\cdot \lbrack e^{-i(\tilde{\omega}\tau _{1}^{\prime
}+\omega \tau _{2}^{\prime })}-e^{-i(\omega \tau _{1}+\tilde{\omega}\tau
_{2})}]d\omega d\tilde{\omega}
\end{equation}%
with
\begin{equation}
F=F(\omega ,\tilde{\omega})=\frac{1}{2}g(\omega )f(\omega ,\tilde{\omega})g(%
\tilde{\omega}).
\end{equation}

\section{The necessary-sufficient condition for the destructive coincidence}

Now let's prove that the necessary-sufficient condition for the destructive
coincidence $R_{c}=0$ at $D=0$ is $f(\omega ,\tilde{\omega})=$ $f(\tilde{%
\omega},\omega )$. To show the sufficience we decompose the two-photon wave
packet $\Psi $ $=\Psi _{1}+\Psi _{2}$ into two parts%
\begin{eqnarray}
\Psi _{1} &=&\int_{0}^{\infty }d\omega d\tilde{\omega}F\cdot e^{-i(\omega
\tau _{1}^{\prime }+\tilde{\omega}\tau _{2}^{\prime })}[1-e^{-i(\omega -%
\tilde{\omega})\frac{D}{C}}] \\
\Psi _{2} &=&\int_{0}^{\infty }d\omega d\tilde{\omega}G(\omega ,\tilde{\omega%
})\cdot e^{-i(\omega \tau _{2}^{\prime }+\tilde{\omega}\tau _{1}^{\prime })}
\nonumber
\end{eqnarray}%
where $G(\omega ,\tilde{\omega})=F(\omega ,\tilde{\omega})-F(\tilde{\omega}%
,\omega )$ in the derivation, we have considered $\tau _{1}-\tau
_{1}^{\prime }=\tau _{2}^{\prime }-\tau _{2}=\frac{D}{C}$. It is easy to see
that $\Psi _{2}=0$ and $\Psi =\Psi _{1}$when $G(\omega ,\tilde{\omega})=0$.
On the other hand, $\Psi _{1}|_{D=0}=0$. Then, at $D=0$, $R_{c}=0.$ To show
the necessity, we consider $|\Psi |^{2}=|\Psi _{1}|^{2}+|\Psi _{2}|^{2}+$ $%
2Re\Psi _{1}^{\ast }\Psi _{2}$. Since $\Psi _{1}=0$ at $D=0$ , $|\Psi
|^{2}|_{D=0}=|\Psi _{2}|^{2}$. So if $R_{c}|_{D=0}=0$, then $|\Psi
|^{2}|_{D=0}=|\Psi _{2}|^{2}=0$ i.e., $\Psi _{2}=0$. Because $G(\omega ,%
\tilde{\omega})$ is the Fourier component of $\Psi _{2}$, the vanishing $%
\Psi _{2}$ must lead to $G=0$ or $f(\omega ,\tilde{\omega})=f(\tilde{\omega}%
,\omega )$.

In fact this necessary-sufficient condition proved above has been implied in
several discussions [19-22].

To illustrate the main idea of \ physics implied by the above proof, we
consider the simplest symmetric state $|\Psi \rangle =|S(\Omega ,\tilde{%
\Omega})\rangle $ $\equiv $ $\frac{1}{\sqrt{2}}\left( |\Omega ,\tilde{\Omega}%
\rangle +|\tilde{\Omega},\Omega \rangle \right) ,$which is an entangled
state in frequency domain. Here, $|\Omega ,\tilde{\Omega}\rangle
=a_{A}^{+}(\Omega )a_{B}^{+}(\tilde{\Omega})|0\rangle .$ The corresponding
two photon wave packet $\Psi _{\Omega \tilde{\Omega}}=$ $\frac{1}{2\sqrt{2}}%
\left( \phi _{\Omega \tilde{\Omega}}+\phi _{\tilde{\Omega}\Omega }\right)
g\left( \Omega \right) g\left( \tilde{\Omega}\right) $ has two parts $\phi
_{\Omega \tilde{\Omega}}=$ $e^{-i\left( \tilde{\Omega}\tau _{1}^{\prime
}+\Omega \tau _{2}^{\prime }\right) }-e^{-i\left( \Omega \tau _{1}+\tilde{%
\Omega}\tau _{2}\right) }$ and $\phi _{\tilde{\Omega}\Omega }$ times the
state density of the light field. Notice that the part $\phi _{\Omega \tilde{%
\Omega}}$ is contributed by the component $|\Omega ,\tilde{\Omega}\rangle $
and $\phi _{\tilde{\Omega}\Omega }$ by $|\tilde{\Omega},\Omega \rangle $. A
straightforward calculation shows that $\Psi _{\Omega \tilde{\Omega}}=0$ at $%
D=0$. We think this phenomenon can not be understood as \textquotedblright
interference between two photons\textquotedblright . The reason \ is, at $%
D=0,$ neither $\phi _{\Omega \tilde{\Omega}}$ nor $\phi _{\tilde{\Omega}%
\Omega }$ is zero, but the sum of them is zero. In fact, the single term $%
\phi _{\Omega \tilde{\Omega}}$ ( $\phi _{\tilde{\Omega}\Omega }$ )
corresponds to the scattering of two independent photons in $|\Omega ,\tilde{%
\Omega}\rangle $ ( $|\tilde{\Omega},\Omega \rangle $) on the beam splitter.
Due to their different frequencies $\tilde{\Omega}$ and $\Omega $, two
independent photons can not interfere with each other. In this sense the
destructive interference can only be attributed to the fact that
\textquotedblright\ the two-photon interferes with itself\textquotedblright
, which is similar to Dirac's statement for single photon.

Since any symmetric two photon state $|\Psi \rangle $ can be decomposed in
general as a "sum" of many symmetric basis vectors $|S(\omega ,\tilde{\omega}%
)\rangle ,$the above observation means that the destructive phenomenon in
two photon coincidence is caused by the symmetric components $|S(\omega ,%
\tilde{\omega})\rangle $ of the light field. Correspondingly the two wave
packet $\Psi $ is a "double sum" of $\Psi _{\omega \tilde{\omega}}$ over $%
(\omega ,\tilde{\omega})$. As $D=0$, every $\Psi _{\omega \tilde{\omega}}$
approaches to zero and then $\Psi $ is also zero. Since the null phenomenon $%
\Psi |_{D=0}=0$ was contributed by every component $\Psi _{\omega \tilde{%
\omega}}|_{D=0}=0$, we are led to the universal conclusion that there does
not exist an effect of \textquotedblright interference between the two
photons\textquotedblright\ . This conclusion holds for the separable case
with a symmetric factorized distribution function $f(\omega ,\tilde{\omega}%
)=f_{1}(\omega )f_{1}(\tilde{\omega}),$ namely, the two independent pulses
have the same shape exactly. In this case this destructive interference
phenomenon of two photon coincidence may occur due to the pairing mechanism
just mentioned above.

\section{\protect\bigskip Free induction decay mechanism for destructive
interference}

In the remaining part of this letter, we will use the free induction decay
mechanism to explain the anti-exponential decaying behavior beside ''dip''\
(or anti-peak) point. It is easy to see that each component $|S(\Omega ,%
\tilde{\Omega})\rangle $ in the symmetric two photon state contributes the
coincidence counting rate $R_{c}$ with the oscillating term
\begin{equation}
R_{c}(\Omega ,\tilde{\Omega})\propto T^{2}\left[ 1-\cos \left( \Omega -%
\tilde{\Omega}\right) \frac{D}{C}\right]
\end{equation}%
Here, the term oscillating with $T$ has been omitted in the long $T$ limit.
Apparently, in this sense , $R_{c}$ is an oscillation function of $D$ with a
constant frequency $\left( \Omega -\tilde{\Omega}\right) $. Then, $R_{c}$ is
zero not only at $D=0$ , but also at $D=D_{k}=\frac{2k\pi c}{\Omega -\tilde{%
\Omega}}\left( k=\pm 1,\pm 2,...\right) $. The oscillating term $\Psi
_{\omega \tilde{\omega}}$ of different frequencies in the two photon wave
packet can cancel each other beside the points $D=0$ and $D_{k},$ this will
\ explain the ''dip''\ experimental curve of non-oscillation in the two
photon coincidence. Based on this conception of the free induction decay
mechanism [15], the calculation of the coincidence counting rate $R_{c}$
leads to the explicit result for the general symmetric input state
\begin{equation}
R_{c}\propto \int_{-\infty }^{\infty }d\omega d\tilde{\omega}|F|^{2}\{1-\cos
[\left( \omega -\tilde{\omega}\right) \frac{D}{C}]\}
\end{equation}%
Here, the integral domains of variables $\tau _{1}^{\prime }$, $\tau
_{2}^{\prime }$, $\omega $ and $\tilde{\omega}$ have been extended to $%
(-\infty ,+\infty )$ for Gaussian type distributions $f(\omega ,\tilde{\omega%
})$. It is easy to see that each oscillating term of variable $D$ in $R_{c}$
has a frequency $(\omega -\tilde{\omega})$. If $|F|^{2}$ is not very steep
at a certain point $(\omega _{0},\tilde{\omega}_{0}),$the double sum of $%
\cos \left[ \left( \omega -\tilde{\omega}\right) \frac{D}{C}\right] $ with
weight $|F(\omega ,\tilde{\omega})|^{2}$ will become a decaying function of
variable $D$ with the maximum at $D=0$ and then $R_{c}$ will become an
anti-decaying function. This is a typical optical free induction decay
phenomenon for the second order coherence function.

For an illustration of the two photon free induction decay, let's calculate
a separable case with $F(\omega ,\tilde{\omega})=F(\omega )F(\tilde{\omega})$
where $F(\omega )$ is of Gaussian type, i.e., $F(\omega )=\exp [-\frac{%
\left( \omega -\omega _{0}\right) ^{2}}{2\sigma ^{2}}]$. It is
straightforward to calculate $R_{c}$ , obtaining the result $R_{c}\propto
\left( 1-e^{-\frac{D^{2}}{2C^{2}}\sigma ^{2}}\right) $. From this
calculation we observe that, even for a factorizable two-photon input state,
there still occurs the destructive phenomenon $R_{c}|_{D=0}=0$ with the
anti-exponential Gaussian decay as $|D|$ increases if the two photons are
created at the same time and their shapes are strictly the same i.e ., $%
f(\omega )=f^{\prime }(\omega )$. It is remarkable that this phenomenon i.e.
the interference of two independent photon has been analized theoretically
and realized experimentally recently [23,24].

To investigate the possibility of testing the present theoretical
predictions, we have consider the practical cases where the condition $%
G(\omega ,\tilde{\omega})=0$ is not strictly satisfied. Obviously, just at $%
D=0,$ the counting rate of two photon coincidence%
\begin{equation}
R_{c}|_{D=0}=\int_{-\infty }^{\infty }d\omega d\tilde{\omega}|G(\omega ,%
\tilde{\omega})|^{2}
\end{equation}%
measures the average extent of asymmetry of the input two photon states,
which depends on the difference between $F(\omega ,\tilde{\omega})$ and $F(%
\tilde{\omega},\omega )$. On the other hand, around the point $D=0$,
\begin{eqnarray}
R_{c}  =  \int |\Psi _{1}|^{2}d\tau _{1}^{\prime }d\tau _{2}^{\prime
}+\int |\Psi _{2}|^{2}d\tau _{1}^{\prime }d\tau _{2}^{\prime }\\
& &+2Re\int \Psi _{1}^{\ast }\Psi _{2}d\tau _{1}^{\prime }d\tau
_{2}^{\prime } \nonumber
\end{eqnarray}%
It is easy to prove that, as $|D|$ increases, the first term in $R_{c}$ will
increase while the second term keeps a constant independent of $D$. However,
the presence of the third term may suppress the increasing tendency of $%
R_{c} $ as $|D|$ increases. In the case that $|G(\omega ,\tilde{\omega})|$is
very small and $|\Psi _{2}|<<|\Psi _{1}|$, the shape of $R_{c}$ in the
neighbor around $D=0$ is just similar to that for the ideal case with $%
G(\omega ,\tilde{\omega})=0$.

In Fig3, we give the curve of "dip" for the rate of two photon coincidence
in a typical non-symmetric case that $G(\omega ,\tilde{\omega})\neq 0$ and $%
F(\omega ,\tilde{\omega})=F_{1}(\omega )F_{2}(\tilde{\omega})$ where $%
F_{1}(\omega )=\exp (-\frac{(\omega -\Omega )^{2}}{2\sigma ^{2}})$ and $%
F_{2}(\tilde{\omega})=\exp (-\frac{(\tilde{\omega}-\tilde{\Omega})^{2}}{%
2\sigma ^{2}})$. The asymmetry parameter $y$ is defined by $y=1-\frac{\tilde{%
\Omega}}{\Omega }$. It describes the relative difference between $\Omega $
and $\tilde{\Omega}$. In this case we have $R_{c}\propto 1-\exp (-\frac{%
\sigma ^{2}D^{2}}{2c^{2}}-\frac{\omega ^{2}y^{2}}{2\sigma ^{2}})$. It is
easy to see that, as $y$ increases, the value of $R_{c}|_{D=0}$ increases
and " the $R_{c}-$ curve" becomes more and more flat.

%
\begin{figure}[h]
\begin{center}
\includegraphics[width=5cm,height=5cm]{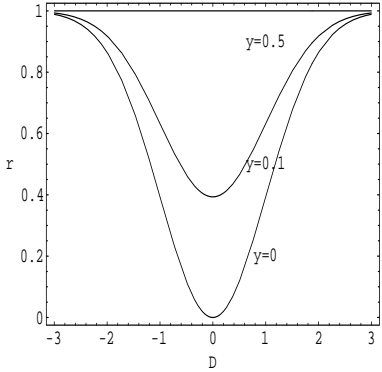}
\end{center}
\caption{The curve of $R_c=r$ as a function of $D$ when $F(\protect\omega ,%
\tilde{\protect\omega})=e^{-\frac{(\protect\omega-\Omega)^2}{2\protect\sigma%
^2}}e^{-\frac{(\tilde{\protect\omega}-\tilde{\Omega})^2}{2\protect\sigma^2}}$%
. Here we assume $\frac {\Omega}{\protect\sigma}=10$. The unit of $D$ is $%
\frac c{\protect\sigma}$.}
\end{figure}

Another more interesting illustration is that $F(\omega ,\tilde{\omega}%
)=F_{1}(\omega )F_{1}(\tilde{\omega})e^{-i\omega T}$. This distribution
function means that the two-photon wave packet can be factorized as the
product of two signal photon wave packets of the same shape, one (in mode $%
A) $of which was created earlier than another (in mode $B)$ with time $T$.
In this case, it is easy to prove that $R_{c}(T,D)=R_{c}(0,D+cT)$ . In Fig.3
.

%
\begin{figure}[h]
\begin{center}
\includegraphics[width=5cm,height=5cm]{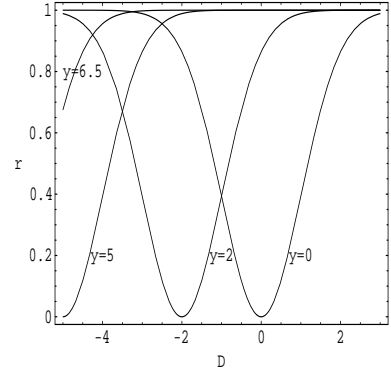}
\end{center}
\caption{The curve of $R_c=r$ as a function of $D$ when $F(\protect\omega ,%
\tilde{\protect\omega})=e^{-\frac{(\protect\omega-\Omega)^2}{2\protect\sigma%
^2}}e^{-\frac{(\tilde{\protect\omega}-\protect\omega)^2}{2\protect\sigma^2}%
}e^{-i\protect\omega T }$. Here we define parameter $y=T\protect\sigma$. The
unit of $D$ is $\frac c{\protect\sigma}$.}
\end{figure}

So far we have only discussed the ideal situation with $50-50$ beamsplitter
where the reflectivity ($R$) and transmissivity ($T$) of the beam splitter
are all $1/2$. In the general case $R\neq T$, we have
\begin{eqnarray}
E_{1}^{+}(t_{1}) &= & i\sqrt{R}E_{a}^{+}(\tau _{1})+\sqrt{T}E_{b}^{+}(\tau
_{1}^{\prime }) \\
E_{2}^{+}(t_{2}) & = & \sqrt{T}E_{a}^{+}(\tau _{2}^{\prime })+i\sqrt{R}%
E_{b}^{+}(\tau _{2}) \nonumber%
\end{eqnarray}%
and the corresponding two-photon wave packet$\Psi (t_{1},t_{2}).$In this
general case, the necessary-sufficient condition for $R_{c}|_{D=0}=0$ can
never be satisfied unless $R=T=1/2$. In the experiment [9] where $G\left(
\omega ,\tilde{\omega}\right) =0$ and $R\neq T$, the effective wave packet $%
\Psi $ can also be decomposed into two parts:$\Psi _{1}\propto
R\int_{0}^{\infty }d\omega d\tilde{\omega}(...)$ $\times (1-e^{-i(\omega -%
\tilde{\omega})\frac{D}{C}}),\Psi _{2}\propto (T-R).$Apparently, when $D=0$,
$\Psi _{1}=0$ and $R_{c}|_{D=0}=|\Psi _{2}|^{2}$. Therefore, $\Psi _{2}$
characterizes the basic difference between $R$ and $T$. If $|\frac{R}{T}-1|$
is very small, $\Psi _{2}$ can be considered as a perturbation for $\Psi
_{1} $.

\section{\protect\bigskip Two photon interference of Type II}

The above analysis can be generalized to the case of photon's polarization.
For instance, we can consider the two-photon interference demonstrated in
Fig.5. 
\begin{figure}[h]
\begin{center}
\includegraphics[width=7cm,height=4cm]{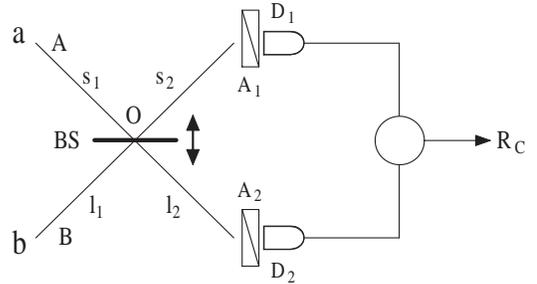}
\end{center}
\caption{Schematic of the two-photon interferometer with polarization
considered}
\end{figure}
In this theoretical scheme of two-photon interference experiment, a photon
pair state with entanglement with polarization degrees of freedom can be
written as%
\begin{eqnarray}
\left\vert \Psi \right\rangle &=&\int_{0}^{\infty }d\omega d\widetilde{%
\omega }f\left( \omega ,\widetilde{\omega }\right) \times  \label{c} \\
&&\frac{1}{\sqrt{2}}\left[ a_{AH}^{+}\left( \omega \right) a_{BV}^{+}\left(
\widetilde{\omega }\right) -a_{AV}^{+}\left( \omega \right) a_{BH}^{+}\left(
\widetilde{\omega }\right) \right] \left\vert 0\right\rangle \text{.}
\nonumber
\end{eqnarray}%
The photons arrive detector $D_{1}$ and $D_{2}$ after mixed by the beam
splitter $BS$. Here $a_{AH}^{+}\left( \omega \right) $ is the creation
operator of photon of frequency $\omega $, the wave vector is along
direction $Oa$ (in optical path mode $A$) with horizontal polarization
(along direction $\widehat{e}_{H}$). The notations $a_{BV}^{+}\left(
\widetilde{\omega }\right) $, $a_{AV}^{+}\left( \omega \right) $ and $%
a_{BH}^{+}\left( \widetilde{\omega }\right) $ are defined similarly as $%
a_{AH}^{+}\left( \omega \right) .$\ Two polarization analyzers $A_{1}$ and $%
A_{2}$ along the directions $\widehat{A}_{1}$ and $\widehat{A}_{2}$ are put
in front of the detectors.

It is easy to see that the rate $R_{c}$ of coincident detection is
proportional to the time integral of the square of effective wave function $%
\psi $'s mode, i.e.%
\[
R_{c}\propto \int dt_{1}dt_{2}\left\vert \psi \left( t_{1},t_{2}\right)
\right\vert ^{2}
\]%
where
\begin{equation}
\psi \left( t_{1},t_{2}\right) =\left\langle 0\right\vert E_{1}^{+}\left(
t_{1}\right) E_{2}^{+}\left( t_{2}\right) \left\vert \Psi \right\rangle
\text{.}  \label{b}
\end{equation}%
Here $E_{i}^{+}\left( t_{i}\right) =\widehat{A}_{i}\cdot \overrightarrow{%
E_{i}^{+}}\left( t_{i}\right) $,and$\overrightarrow{\text{ }E_{i}^{+}}\left(
t_{i}\right) =\widehat{e}_{H}$ $E_{H}^{+}\left( t_{i}\right) +\widehat{e}%
_{V} $ $E_{V}^{+}\left( t_{i}\right) $( $i=1,2)$ are the positive frequency
parts of light field operator at the position of the detector. Considering
the transforming character of beam splitter, we have
\begin{eqnarray}
\overrightarrow{E_{1}^{+}}(t_{1}) &=&\frac{1}{\sqrt{2}}[i\overrightarrow{%
E_{a}^{+}}(\tau _{1})+\overrightarrow{E_{b}^{+}}(\tau _{1}^{\prime })]
\label{a} \\
\overrightarrow{E_{2}^{+}}(t_{2}) &=&\frac{1}{\sqrt{2}}[\overrightarrow{%
E_{a}^{+}}(\tau _{2}^{\prime })+i\overrightarrow{E_{b}^{+}}(\tau _{2})]\text{%
.}  \nonumber
\end{eqnarray}%
Here $\tau _{1}$, $\tau _{1}^{\prime }$, $\tau _{2}$, $\tau _{2}^{\prime }$
are just defined as in section 2 and $\overrightarrow{E_{a}^{+}}_{\left(
b\right) }=\widehat{e}_{H}$ $E_{a\left( b\right) H}^{+}+\widehat{e}_{V}$ $%
E_{a\left( b\right) V}^{+}$ are the positive frequence part of the light
field operator at point $a$ and $b$ and $E_{a\left( b\right) H\left(
V\right) }^{+}\left( t\right) =\int d\omega a_{A\left( B\right) H\left(
V\right) }\exp \left[ -i\omega t\right] $. The spectral density function $%
g\left( \omega \right) $ is neglected here without loss of generity.

Submitting \ref{a} to \ref{b}, we can get the explicit expression \ for
effective wave function $\psi $ of the light field's $:$%
\begin{eqnarray}
\psi &=&\frac{1}{2\sqrt{2}}\left[ \widehat{A}_{1}\cdot \widehat{e}_{V}\times
\widehat{A}_{2}\cdot \widehat{e}_{H}-\widehat{A}_{1}\cdot \widehat{e}%
_{H}\times \widehat{A}_{2}\cdot \widehat{e}_{V}\right] \times \\
&&\int d\omega d\widetilde{\omega }f\times \left[ e^{-i\left( \widetilde{%
\omega }\tau _{1}^{\prime }+\omega \tau _{2}^{\prime }\right) }+e^{-i\left[
\omega \tau _{1}^{\prime }+\widetilde{\omega }\tau _{2}^{\prime }+\left(
\omega -\widetilde{\omega }\right) \frac{D}{c}\right] }\right]  \nonumber
\end{eqnarray}%
where $D=l_{1}-s_{1}=c\left( \tau _{1}-\tau _{1}^{\prime }\right) $ $%
=c\left( \tau _{2}^{\prime }-\tau _{2}\right) $ is the optical path
difference disscussed before. Then $R_{c}$ can be written as%
\begin{equation}
R_{c}=\xi \left( \widehat{A}_{1},\widehat{A}_{2}\right) \times \eta \left(
D\right)
\end{equation}%
where%
\begin{equation}
\xi \left( \widehat{A}_{1},\widehat{A}_{2}\right) =\frac{1}{8}\left\vert
\widehat{A}_{1}\cdot \widehat{e}_{V}\times \widehat{A}_{2}\cdot \widehat{e}%
_{H}-\widehat{A}_{1}\cdot \widehat{e}_{H}\times \widehat{A}_{2}\cdot
\widehat{e}_{V}\right\vert ^{2}
\end{equation}%
is dependent of the direction of polarization analyser and%
\begin{eqnarray}
\eta \left( D\right) &=&\int d\tau _{1}^{\prime }d\tau _{2}^{\prime }|\int
d\omega d\widetilde{\omega }f\times  \nonumber \\
&&\left[ e^{-i\left( \widetilde{\omega }\tau _{1}^{\prime }+\omega \tau
_{2}^{\prime }\right) }+e^{-i\left[ \omega \tau _{1}^{\prime }+\widetilde{%
\omega }\tau _{2}^{\prime }+\left( \omega -\widetilde{\omega }\right) \frac{D%
}{c}\right] }\right] |^{2}
\end{eqnarray}%
varies as the optical path difference $D$ is changed.

It is easy to prove that, when the frequency distribution function $f$ is
symetric, i.e. $f\left( \omega ,\widetilde{\omega }\right) =f\left( \omega ,%
\widetilde{\omega }\right) ,$ we have $\frac{d\eta }{dD}\mid _{D=0}$and
\begin{equation}
\frac{d^{2}\eta }{dD^{2}}\mid _{D=0}\propto -\int d\omega d\widetilde{%
\omega }\left\vert f\right\vert ^{2}\left( \omega -\widetilde{\omega }%
\right) ^{2}<0.
\end{equation}
So we know that in this case $R_{c}$ has its maximal value when $D=0$. This
is differenct to the case in which the photon polarization is not considered
and $R_{c}$ has its minimal value when $D=0$. In fact, this difference is
due to the $-1$ factor in light field's initial state \ref{c} i.e. if the
initial state is
\begin{equation}
\int_{0}^{\infty }d\omega d\widetilde{\omega }f\times \frac{1}{\sqrt{2}}%
\left[ a_{AH}^{+}\left( \omega \right) a_{BV}^{+}\left( \widetilde{\omega }%
\right) +a_{AV}^{+}\left( \omega \right) a_{BH}^{+}\left( \widetilde{\omega }%
\right) \right] \left\vert 0\right\rangle ,
\end{equation}%
$R_{c}$ will also has its minimal value when $D=0$.

It is notable that when the photon's polarization is considered, $f\left(
\omega ,\widetilde{\omega }\right) =f\left( \omega ,\widetilde{\omega }%
\right) $ is only the sufficient condition of that $R_{c}$ has its maximal
value (or minimal value) when $D=0$ but not the necessary condition. In
fact, it is difficult to obtain its necessary condition.

\bigskip

\section{Concluding remarks}

To summarize, we have theoretically investigated the two photon interference
phenomenon for a general situation with an arbitrary two photon state and
given the necessary-sufficient condition for the destructive interference of
coincidence counting in terms of the symmetric pairing of photons in the
light pulses,i.e.,$G(\omega ,\tilde{\omega})=0$. This implies that, even in
the separable case $f(\omega ,\tilde{\omega})=f_{1}(\omega )f_{2}(\tilde{%
\omega})$, if $f_{1}=f_{2},$a curve of \textquotedblright
dip\textquotedblright\ can still be observed in the rate of two photon
coincidence. Our investigation predicts the destructive interference
phenomenon to occur even for certain cases with separable input two photon
state. However, this does not imply \textquotedblright the interference
between two independent photons\textquotedblright\ since the essence leading
to destructive interference is that the required two photon state consists
of the inseparable symmetric component$|S(\omega ,\tilde{\omega}\rangle .$
Since for the the symmetric distribution function all components of the
light field state have the entangled forms $|S(\omega ,\tilde{\omega}\rangle
$,we understand the \textquotedblright dip\textquotedblright\ in the
destructive coincidence curve according to the free induction decay
mechanism from the dispersion of the two-photon frequency ($\omega ,\tilde{%
\omega}$). In fact, though each component contributes $R_{c}$ with an
oscillating term with respect to $D$, in the integral of variables $\tilde{%
\omega}$ and $\omega $ with Gaussian type distribution $f$ , these
oscillating terms cancel one another and thus lead to a (anti-) Gaussian
decaying factor. We also considered the effect of photon's polarization. We
found that,when the effective wave function is symetric i.e. $f\left( \omega
,\tilde{\omega}\right) =f\left( \tilde{\omega},\omega \right) $, the
coincidence counting rate may have its maximal or minimal value when the
path difference is zero.

\textbf{Acknowledgement:} \textit{The authors thank Jian-wei Pan for his
useful discussions. This work is supported by the NSF of China ( CNSF grant
No.90203018) and the Knowledged Innovation Program(KIP) of the Chinese
Academy of Science. It is also founded by the National Fundamental Research
Program of China with No 001GB309310.}

\end{document}